\documentclass[reprint,amsmath,amssymb,smsymb,aps,superscriptaddress,apl,prd,nofootinbib, floatfix]{revtex4-2}
\usepackage{geometry}
\usepackage{graphicx}
\usepackage{mathtools}

\usepackage{xcolor}
\usepackage[normalem]{ulem}

\begin{document}
\title{Parameter estimation bias from overlapping binary black hole events in second generation interferometers}

\author{Philip Relton}
\email[Correspondence email address: ]{reltonpj@cardiff.ac.uk}
\author{Vivien Raymond}
\affiliation{Gravity Exploration Institute, Cardiff University, Cardiff CF24 3AA, United Kingdom}
\date{\today}

\begin{abstract}
    Since the initial detection of gravitational waves in 2015, over 50 candidate events have been reported by the LIGO-Virgo-KAGRA collaboration. As the current generation of detectors move toward their design sensitivity, the rate of these detections will increase. The next generation of detectors are likely to have high enough sensitivities that multiple merging binaries will be visible at the same time. In this paper we show that this is likely to happen before the end of the decade, during observations by the LIGO-Voyager detector. We investigate the situation of overlapping binary black hole (BBH) mergers in these detectors. We find that current parameter estimation techniques are capable of distinguishing the louder of two merging BBH events, without significant bias, when their merger times are not less than $\sim0.1$ seconds apart or when the ratio of the signal-to-noise ratios of the systems is uneven. This region of overlapping parameter space is dependent upon the sky locations of the signals and the relation of those locations to the light travel time between detectors. We also find that, if two signals are highly overlapping, then the recovered set of parameters often show strong evidence of precession. Finally, we show that bias can occur even when the signal causing the bias is below the detection threshold.
\end{abstract}

\keywords{gravitational waves, parameter estimation, overlapping signals}

\maketitle

\section{INTRODUCTION} \label{intro}

In 2015 the Advanced LIGO (aLIGO) detectors \cite{aasi2015advanced} began their first observing runs \cite{abbott2016binary, abbott2016gw150914}, with the Advanced Virgo detector joining in 2017 \cite{acernese2014advanced}. During approximately a year of observations; a binary neutron star (BNS) merger \cite{abbott2017gw170817} and eleven binary black bole (BBH) mergers were observed \cite{abbott2019gwtc}, at a rate of roughly one per five weeks. After another year of upgrades ~\cite{acernese2019increasing, tse2019quantum, buikema2020sensitivity}, the combined LIGO-Virgo-KAGRA (LVK) Collaboration observed 39 mergers within six months of observing time. This corresponds to a rate of roughly 1.5 mergers per week \cite{abbott2021gwtc}. Candidates for further events have been reported, using LIGO-Virgo data from the first two observing runs, by groups external to the LVK Collaboration ~\cite{nitz20202, venumadhav2020new, zackay2019highly, zackay2021detecting}. As gravitational wave (GW) detectors progress to their design sensitivity \cite{aasi2015advanced}, and are replaced with more advanced observatories \cite{baibhav2019gravitational, adhikari2020cryogenic, hild2010xylophone, punturo2009einstein, reitze2019cosmic}, the viewing range of the detectors will increase. This in turn will increase the rate of binary merger detection.

As the rate of observed mergers increases, the probability of concurrent signal being observed by the network increases. Most current detection and analysis techniques assume that only one signal is visible at any period of time. Therefore, any overlapping signals could be assumed to be one single signal. If overlapping signals start to make up a significant fraction of detected signals, their effects could bias wider reaching studies such as population modelling and tests of general relativity. Even a single overlapping event could negatively affect scientific results if the bias from the second signal causes significant deviation from a true signal.

In this paper we investigate the probability of observing two compact binary mergers at close enough times that their waveforms interfere while observable in the detector. Previous studies ~\cite{pizzati2021bayesian, samajdar2021biases} have performed this calculation numerically and have focused on third generation detectors (3G), such as Einstein Telescope ~\cite{punturo2009einstein, hild2010xylophone} and Cosmic Explorer \cite{reitze2019cosmic}. Here we calculate this analytically, including results for second generation detectors and proposed end of generation detectors such as LIGO-Voyager ~\cite{collaboration2015instrument, baibhav2019gravitational}. 

We also investigate the effects overlapping signals can have on parameter estimation techniques currently in use within the LVK Collaboration. Previous studies into overlapping signals have investigated the bias caused when attempting to recover either of the two signals in the data ~\cite{pizzati2021bayesian, samajdar2021biases}. For our analysis, we treat overlapping signals as a single event and apply parameter estimation in an agnostic manner, making only the assumption that the system is a merging BBH. We attempt to constrain the regions in which such overlapping signals show significant bias. We also show what forms this bias can take when analysis with parameter estimation techniques includes sampling over black hole spin parameters and uses waveforms that include spin-induced precession. See ~\cite{apostolatos1994spin, kidder1995coalescing} for an explanation of spin-induced precession.

We show the calculation of the probability of overlapping observations Sec. \ref{prob}. We then outline our method of signal overlap and recovery in Sec. \ref{method}. The results of this study are described in Sec. \ref{results}, with updates on our probability calculation from our results in Sec. \ref{upprob}. Sec. \ref{conclusions} provides a summary of our findings and proposals of avenues for further study.

\section{PROBABILITY} \label{prob}
\subsection{Probability of observing overlapping events} \label{probOL}
The number of compact binary coalescence (CBC) events observed by a detector depends on the merger rate of such systems, the length of the observing period and the volume of space in which that merger is visible by the detector. It is possible to characterise the noise of a detector by taking a power spectral density (PSD). This can further be used to define the signal-to-noise ratio (SNR) of any signal, including modelled compact binary coalescence, see Chap. 7 of \cite{creighton2012gravitational} for detailed derivations of the SNR and explanations of PSDs.

As CBC merger times are assumed to be random and independent, they can be modelled by a Poisson distribution. The probability of a merging system being observed within a specific time period is then given by:

\begin{equation}\label{poisson}
    P(\mathrm{k\:events\:in\:time\:T_{obs}}) = \frac{(R\,T_{obs})^{k}\exp^{-RT_{obs}}}{k!}
\end{equation}
where $R$ is the rate of mergers, $T_{obs}$ is the period of observation, $k$ is the number of mergers in that period. For two mergers to overlap in time, there must be two events that occur within the period of a single signal, $T_{signal}$. With this period we can use Poisson statistics to calculate the inter-arrival time for two events. If the first event occurs at time $t_0$ and the second event occurs at time $t_1 = t_0 + \Delta T$, then we can find the probability that $\Delta T$ is less than the observable period of the signal:

\begin{equation}\label{interarrival}
    P(\Delta T\,<\,T_{signal}) = 1 - \exp^{-RT_{signal}}
\end{equation}

This is the probability that a detected signal is overlapping another signal. Further, by estimating the number of detectable signals in an observing period of a detector; an estimate as to the number of overlapping signals can be found. This rough estimate is simply the number of detectable events multiplied by the probability of events overlapping.

Current estimates on the merger rate of compact binaries are confined to redshifts of $z \lesssim 1$ due to the nature of observed events. It is possible that this rate will increase with increasing redshift \cite{abbott2021population}. An increase in the rate of detections would lead to an increase in the rate of overlapping signals. However, for the purposes of our study we have assumed a constant merger rate, regardless of redshift. This assumption should hold well for all detectors discussed here, with the exception of Einstein Telescope, as their viewing ranges are constrained within the volume $z \lesssim 1$. Calculated values for Einstein Telescope should be treated with caution due to this uncertainty.

To estimate the distance at which a particular detector configuration can observe different merging systems we use the BNS range. By convention, the BNS range of a detector is the distance at which the detector can observe a binary neutron star coalescence with masses $1.4 \mathrm{M_{\odot}} - 1.4 \mathrm{M_{\odot}}$ at an SNR of 8. The equivalent for binary black hole mergers is for two $30\,\mathrm{M_{\odot}}$ black holes at an SNR of 8. The mean BNS range for the first part of the third observing run, O3, were $108\,\mathrm{Mpc}$ and $135\,\mathrm{Mpc}$ for LIGO: Hanford and LIGO: Livingston respectively \cite{abbott2021gwtc}. Estimates of the BBH range, prior to the start of O3, indicate it should have been in the range $990 - 1200\,\mathrm{Mpc}$ \cite{abbott2020prospects}.

\subsection{Probability of overlapping BBH mergers} \label{P_BBH_OL}
The rates and populations paper from the second gravitational wave transient catalogue produced by the LVK collaboration estimated the rate of BBH mergers from the events it observed at: $\mathcal{R}_{\mathrm{BBH}}\,=\,23.9^{+14.9}_{-8.6}\,\mathrm{Gpc}^{-3}\mathrm{yr}^{-1}$\cite{abbott2021population}.

The large sample of observed BBH mergers allows for the rate of merger to be calculated as a function of the systems component masses \cite{abbott2021population}. We used the publicly available data, see \cite{ligo2020randp}, for the Power Law + Peak mass distribution shown in \cite{abbott2021population}. This distribution gives a rate estimate for a variety of different merging black hole systems with primary masses between $[2,100] \mathrm{M_{\odot}}$ and mass ratios between $[0.1,1]$. For each system in this grid we used the \texttt{inspiral-range} package ~\cite{chen2021distance, gwinc-inspiralrange}, alongside estimates of the PSD for different detector configurations ~\cite{aligoO3PSD, aligoO4PSD, aligoPSD, ETPSD, hild2011sensitivity}\footnote{The PSD used for aLIGO: O3 is taken from the first three months of the Livingston detector during O3.}\footnote{The LIGO-Voyager PSD is a noise estimate calculated via the \texttt{pygwinc} python package as described in Sec. \ref{voyager_network}} to estimate the distance at which the detector could observe a merging BBH system of that mass pairing to an SNR of 8. Estimates of the visible duration of the signal were then drawn by generating waveforms for each mass pairing and calculating the duration of the signal between a low frequency cutoff and the merger. These were $20$ Hz for the aLIGO configurations \cite{abbott2020prospects}, $10$ Hz for LIGO-Voyager \cite{collaboration2015instrument} and $1$ Hz for Einstein Telescope \cite{hild2010xylophone}.

Estimates for the probability of an observed signal containing an overlap could then be calculated in the following way:
\begin{equation}\label{poverlap}
    \begin{split}
        P(\mathrm{Overlap}) & = 1 - \exp^{-<RT>} \\
                            & = 1 - \exp^{-<\mathcal{R}V(m)T(m)>} \\
                            & = 1 - \exp^{-\mathcal{R}\int V(m)T(m)p(m)dm}
    \end{split}
\end{equation}
where $\mathrm{<RT>}$ is the average of the merger rate and visible signal duration over all mass pairings, $V(m)$ is the visible volume, calculated as a sphere at inspiral range for mass pairing $m$, $T(m)$ is the visible duration for the mass pairing, and $p(m)$ is the probability of observing the mass pairing.

From this we found the probability that an observed binary, in the next aLIGO operating run, overlaps another signal at approximately ${9.0}^{+5.6}_{-3.2} \times 10^{-6}$. Estimating the number of events, by multiplying the rate of BBH mergers with $VT$, the visible volume and observing duration, leads to an expected $130.0^{+79.0}_{-47.0}$ events in O4. From these values there is very little chance of observing any overlapping events in this observing run. The same is true for the final aLIGO configuration at design specifications.

However, when aLIGO is replaced by the proposed LIGO-Voyager detector ~\cite{baibhav2019gravitational, adhikari2020cryogenic} the sensitivity, and therefore viewing range, will increase. LIGO-Voyager is expected to have a low frequency cutoff of $10\,\mathrm{Hz}$. This increases the observable period for BBH signals, further increasing the probability of overlap.

With this detector configuration, the number of mergers in a year's observing run is $2800.0^{+1800.0}_{-1000.0}$. The probability of observing overlapping signals in LIGO-Voyager is approximately ${1.9}^{+1.2}_{-0.7} \times 10^{-3}$. This could occur in as many as $5.3^{+8.6}_{-3.1}$ of observable BBH detections in LIGO-Voyager per year. This is a significant fraction of events, particularly if the bias from the overlap causes drastic changes to the recovered signal parameters. LIGO-Voyager is therefore the most likely detector to first observe such events.

By the mid 2030s, 3G detectors, such as Einstein Telescope (ET), will start their observation runs. These detectors will have almost an order of magnitude better sensitivity and much lower frequency cutoffs. Observing overlapping signals here is near certain and should account for almost all BBH mergers. Values for these calculations are given in Table \ref{tab:BBH_prob}.

\begin{table*}
    \centering
    \begin{tabular}{|c|c|c|c|c|c|c|c|c|c|}
        \hline
        Detector & Mean BBH & Low Frequency & Mean Visible & $P(Overlap)$ & $N_{events}$ & $N_{Overlap}$ \\
        Configuration & Range (Mpc) & Cut Off (Hz) & Duration (s) & (BBH) & (BBH) & (BBH) \\
        \hline
        aLIGO: O3 & 679.7 & 20 & 6.403 & ${3.5}^{+2.2}_{-1.3} \times 10^{-6}$ & $54.0^{+33.0}_{-19.0}$ & $0.0^{+0.0}_{-0.0}$ \\
        \hline
        aLIGO: O4 & 929.1 & 20 & 6.403 & ${9.0}^{+5.6}_{-3.2} \times 10^{-6}$ & $130.0^{+79.0}_{-47.0}$ & $0.0^{+0.0}_{-0.0}$ \\
        \hline
        aLIGO: Design & 974.8 & 20 & 6.403 & ${1.0}^{+0.6}_{-0.4} \times 10^{-5}$ & $150.0^{+94.0}_{-54.0}$ & $0.0^{+0.0}_{-0.0}$ \\
        \hline
        LIGO-Voyager & 2841.0 & 10 & 40.95 & ${1.9}^{+1.2}_{-0.7} \times 10^{-3}$ & $2800.0^{+1800.0}_{-1000.0}$ & $5.3^{+8.6}_{-3.1}$ \\
        \hline
        Einstein Telescope & 5041.0 & 1 & 18850.0 & $1.0^{+0.0}_{-0.0}$ & $13000.0^{+8500.0}_{-4500.0}$ & $13000.0^{+8500.0}_{-4700.0}$ \\
        \hline
    \end{tabular}
    \caption{The probabilities of detecting overlapping BBH signals, and estimates on the number of such detections, in different GW detector configurations, across a years observations.}
    \label{tab:BBH_prob}
\end{table*}

\subsection{Probability of overlapping BNS mergers} \label{P_BNS_OL}

\begin{figure}[t!]
\centering
\includegraphics[width=0.45\textwidth]{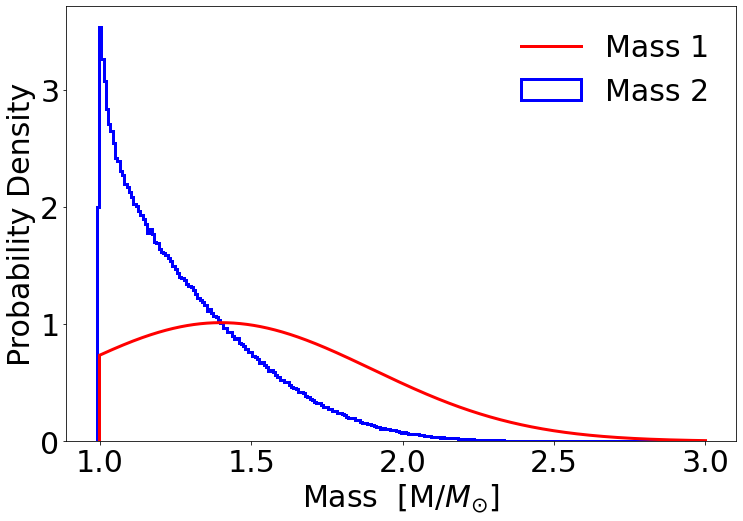}
\caption{Distribution of primary mass, red, secondary mass, blue, for BNS mergers}
\label{fig:BNSmassdist}
\end{figure}

The BNS merger rate is not well constrained due to small number of observed BNS mergers. To date the LVK collaboration has only observed two BNS mergers \cite{abbott2017gw170817, abbott2020gw190412}. To calculate the average period of their merger we defined a probability distribution function for the mass of the primary neutron star and drew a sample grid. The grid covered the range $[1,3]\,\mathrm{M_{\odot}}$ for primary mass and $[0.3,1]$ for mass ratio. For each mass pair the probability of the primary mass was drawn from a truncated normal distribution between $1\,\mathrm{M_{\odot}}$ and $3\,\mathrm{M_{\odot}}$ with a mean of $1.4\,\mathrm{M_{\odot}}$ and variance 0.5. The probability of the secondary mass was drawn from the same distribution but with truncated limits between $1\,\mathrm{M_{\odot}}$ and the primary mass. The probability of that mass pair is then the product of these two probabilities. These two distributions can be seen in Fig. \ref{fig:BNSmassdist}.

We followed the same method described in Sec. \ref{P_BBH_OL} to calculate the probability of two or more overlapping signals occurring in O4. The rate of BNS mergers is $\mathcal{R}_{\mathrm{BNS}}\,=\,320^{+490}_{-240}\,\mathrm{Gpc}^{-3}\mathrm{yr}^{-1}$ \cite{abbott2021population}, the predicted BNS range is $190\,\mathrm{Mpc}$ \cite{abbott2020prospects}. BNS signals have much lower mass than BBH systems and as such merge at much higher frequencies with visible duration on the order of minutes.

The predicted number of observed BNS mergers in O4 is $9.6^{+15.0}_{-7.2}$ mergers per year. Over the full year of O4 the probability of detecting two signals from BNS events that overlap is ${4.2}^{+6.4}_{-3.1} \times 10^{-5}$. Therefore it is not expected that overlapping BNS signals will be observed in the next aLIGO observing run.

The increase in visible volume allowed by LIGO-Voyager, coupled with the increase in visible duration via the lower low frequency cutoff , leads to a drastic increase in the number of observed overlapping BNS mergers. This could account for $\sim 1.7 \%$ of BNS mergers every year. However, our initial studies of overlapping binaries led us to study signals that overlap closely in time periods. This calculation only sets the requirement that the signals overlap in the time domain. Despite the large number of overlapping BNS events in LIGO-Voyager, it is unlikely that any will overlap across a significant portion of the inspiral. Several studies have looked parameter estimation upon such overlaps \cite{pizzati2021bayesian, samajdar2021biases}.

As with BBH mergers, during third generation detector observing runs, it is likely that most, if not all BNS mergers will overlap. This is mostly due to the increased number of observable cycles. Values for these calculations are given in Table \ref{tab:BNS_prob}.

\begin{table*}
    \centering
    \begin{tabular}{|c|c|c|c|c|c|c|c|c|c|}
        \hline
        Detector & Mean BNS & Low Frequency & Mean Visible & $P(Overlap)$ & $N_{events}$ & $N_{Overlap}$ \\
        Configuration & Range (Mpc) & Cut Off (Hz) & Duration (s) & (BNS) & (BNS) & (BNS) \\
        \hline
        aLIGO: O3 & 133.8 & 20 & 151.3 & ${1.5}^{+2.2}_{-1.1} \times 10^{-5}$ & $3.4^{+5.2}_{-2.6}$ & $0.0^{+0.0}_{-0.0}$ \\
        \hline
        aLIGO: O4 & 189.8 & 20 & 151.3 & ${4.2}^{+6.4}_{-3.1} \times 10^{-5}$ & $9.6^{+15.0}_{-7.2}$ & $0.0^{+0.0}_{-0.0}$ \\
        \hline
        aLIGO: Design & 197.7 & 20 & 151.3 & ${4.7}^{+7.2}_{-3.5} \times 10^{-5}$ & $11.0^{+17.0}_{-8.3}$ & $0.0^{+0.0}_{-0.0}$ \\
        \hline
        LIGO-Voyager & 771.6 & 10 & 957.0 & ${1.8}^{+2.6}_{-1.3} \times 10^{-2}$ & $640.0^{+990.0}_{-480.0}$ & $11.0^{+60.0}_{-10.0}$ \\
        \hline
        Einstein Telescope & 2185.0 & 1 & 440000.0 & $1.0^{+0.0}_{-0.0}$ & $14000.0^{+22000.0}_{-10000.0}$ & $14000.0^{+22000.0}_{-10000.0}$ \\
        \hline
    \end{tabular}
    \caption{The probabilities of detecting overlapping BNS signals, and estimates on the number of such detections, in different GW detector configurations, across a years observations.}
    \label{tab:BNS_prob}
\end{table*}

\subsection{Probability of BBH mergers overlapping with BNS mergers} \label{P_BBHBNS_OL}
BBH mergers visible in ground based interferometers for much shorter periods. The visible period for a typical $30+30\,\mathrm{M_{\odot}}$ BBH merger in aLIGO is approximately $0.894$ seconds, a standard $1.4-1.4\,\mathrm{M_{\odot}}$ BNS merger this is $160.8$ seconds. However, at the sensitivities of aLIGO and LIGO-Voyager, BBH signals are observed at a much higher rate. It is therefore possible that overlapping events will be observed with a BBH merger occurring within the visible period of a BNS inspiral.

Here we use the observing range for BBH mergers to estimated the number of BBH mergers observed in the given observing run. To estimate this, we treat the signal as having the period of a BNS merger to account for the overlap with a BNS. As the visible volumes and visible durations now do not depend on the same mass parameter we must modify Eq. \ref{poverlap} to account for the difference:

\begin{equation}\label{poverlap2}
    \begin{split}
        P(\mathrm{Overlap}) & = 1 - \exp^{-<R_{BBH}><T_{BNS}>} \\
                            & = 1 - \exp^{-\mathcal{R}<V(m_{BBH})><T(m_{BNS})>} \\
    \end{split}
\end{equation}

It is possible that such events would be observed before the end of the current aLIGO detectors. They are likely to account for around $8.2\%$ of all BBH mergers in LIGO-Voyager, as many as $230.0^{+360.0}_{-130.0}$ BBH events overlapping BNS signals in a years observation. Values for these calculations are given in Table \ref{tab:BNS+BBH_prob}.

However, our preliminary analysis showed that there should not be significant problems in distinguishing these signals in the case of an overlap. There are large differences between the frequency evolution's of these kinds of CBC mergers. These differences, particularly as most of these BBH signals will merge long before the BNS signal merges, should cause the two events to be distinguishable. This is explained further in Sec. \ref{singledetector}. The portion of the signal containing the merging BBH can be rejected as has been done for glitches present in BNS signals \cite{abbott2017gw170817}. Due to this, we do not present any analysis into these types of overlap here.

\begin{table*}
    \centering
    \begin{tabular}{|c|c|c|c|c|c|c|c|c|c|}
        \hline
        Detector & Mean BNS & Low Frequency & Mean Visible & $P(Overlap)$ & $N_{events}$ & $N_{Overlap}$ \\
        Configuration & Range (Mpc) & Cut Off (Hz) & Duration (s) & (BNS+BBH) & (BBH) & (BNS+BBH) \\
        \hline
        aLIGO: O3 & 679.7 & 20 & 151.3 & ${2.6}^{+1.6}_{-0.9} \times 10^{-4}$ & $54.0^{+33.0}_{-19.0}$ & $0.01^{+0.03}_{-0.0}$ \\
        \hline
        aLIGO: O4 & 929.1 & 20 & 151.3 & ${6.2}^{+3.9}_{-2.2} \times 10^{-4}$ & $130.0^{+79.0}_{-47.0}$ & $0.08^{+0.13}_{-0.05}$ \\
        \hline
        aLIGO: Design & 974.8 & 20 & 151.3 & ${7.2}^{+4.5}_{-2.6} \times 10^{-4}$ & $150.0^{+94.0}_{-54.0}$ & $0.11^{+0.18}_{-0.07}$ \\
        \hline
        LIGO-Voyager & 2841.0 & 10 & 957.0 & ${8.2}^{+4.8}_{-2.9} \times 10^{-2}$ & $2800.0^{+1800.0}_{-1000.0}$ & $230.0^{+360.0}_{-130.0}$ \\
        \hline
        Einstein Telescope & 5041.0 & 1 & 440000.0 & $1.0^{+0.0}_{-0.0}$ & $13000.0^{+8500.0}_{-4500.0}$ & $13000.0^{+8500.0}_{-4500.0}$ \\
        \hline
    \end{tabular}
    \caption{The probabilities of detecting a BBH signal while a BNS signal is present in the detector, and estimates on the number of such detections, in different GW detector configurations, across a years observations. Here the range, and number of individual events, is given for BBH mergers. The duration is for BNS events.}
    \label{tab:BNS+BBH_prob}
\end{table*}

\subsection{Other mergers} \label{Other_OL_events}
The Advanced LIGO and Advanced Virgo detectors are sensitive to several other gravitational wave sources. Since the writing of this work, the mergers of two neutron star-black hole binaries have been observed by the LVK \cite{abbott2021observation}. To the same extent very high mass collisions, such as intermediate mass black hole mergers (IMBH), are rare enough to make their merger rates difficult to predict. These signals are also much shorter, further reducing the chances of signal overlaps. For these reasons we do not consider these events any further in this study.

At the time of writing, no GW event from observed supernovae has been announced by the LVK collaboration. It is of course possible that these events could overlap. However, we do not consider these due to lack of any observations.

Considering the large number of events we expect to see in Einstein Telescope, we examine the probability of observing more than two signals at the same time. This is possible by extending Eq. \ref{interarrival} to the case of $N$ overlapping events:

\begin{equation} \label{eq:prob_multiple}
    P(k > N) = 1 - \sum_{k=0}^{k=N} P(k) = 1 - \sum_{k=0}^{k=N} \frac{(RT_{obs})^{k}\exp^{-RT_{obs}}}{k!}
\end{equation}

\begin{figure}[t!]
\centering
\includegraphics[width=0.45\textwidth]{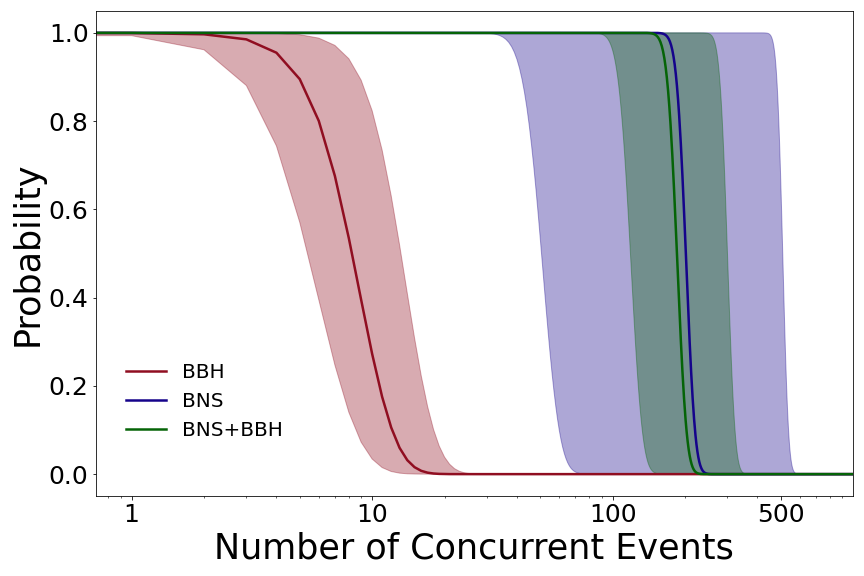}
\caption{Probability distribution for the likely number of BNS, blue, and BBH, red, mergers, that overlap that each event of that type in Einstein Telescope. The likely number of BBH events overlapping each BNS event is also presented in the green curve.}
\label{fig:multi-overlaps}
\end{figure}

We calculated the probability for every value of N up to $1,000$ for three kinds of events; BBH, BNS and BNS+BBH, using the rate and viewing range estimates given in Tables \ref{tab:BBH_prob}, \ref{tab:BNS_prob}, \ref{tab:BNS+BBH_prob} respectively. On average each BBH merger in Einstein Telescope will overlap $8^{+10}_{-6}$ other BBH events. For BNS mergers this is $201^{+45}_{-41}$. These results are shown in Fig. \ref{fig:multi-overlaps}.

It follows from these findings that it is unlikely Einstein Telescope will be able to observe any signal that does not overlap multiple other events. However, the probability of observing $P(k>2)$ in second generation detectors such as aLIGO or LIGO-Voyager is negligible. Therefore, it is reasonable to consider the two signal situation for these detectors.

Our results in Sec. \ref{results} constrain the regions of parameter space in which two signals cause significant bias. In Sec. \ref{upprob} we re-estimate the number of overlapping signals based upon this bias-causing parameter space and apply this to LIGO-Voyager and Einstein Telescope.

\subsection{Validation of analytical probability calculation} \label{Validation}
Previous studies of the probability of observing overlapping signals have considered a similar analytic approach to this calculation \cite{pizzati2021bayesian}. Others have considered a more numerical approach \cite{samajdar2021biases}. For completeness, we here check our result numerically in a similar method to that described in Sec. 2 of \cite{samajdar2021biases}.

From the same distributions as described in Secs. \ref{P_BBH_OL} and \ref{P_BNS_OL} we drew random samples of component masses as estimates of possible observing runs. For each event we calculate the observable period and assigned a merger time drawn from a uniform distribution across a years observing period. We then counted how many signals in that observing period overlap in time. This process was repeated $10^6$ times to obtain a reasonable average for each run. The resulting number of overlaps and P(overlap) agree qualitatively with our analytical predictions. We can therefore conclude that the numbers from our analytical study are accurate, given current rate estimates.

\section{METHOD} \label{method}
\subsection{Parameter estimation of gravitational waves} \label{detection_PE}
The primary signal analysis technique performed in GW analysis is parameter estimation (PE). Typically, parameter estimation uses stochastic sampling to select the set of parameters that best describe the CBC merger. At each step of the parameter estimation the sampler selects a random sample of probable parameter values. The posterior probability, $P(h(t)|d(t))$, of such a signal being present in the data is then calculated via Bayes theorem:

\begin{equation} \label{Bayes}
    P(h(t)|d(t)) = \frac{P(h(t))P(d(t)|h(t))}{P(d(t))}
\end{equation}
where $P(h(t))$ is the prior probability that a signal, $h(t)$ could exist. $P(d(t))$ is the evidence, the probability distribution of the data, $d(t)$. $P(d(t)|h(t))$ is the likelihood, the probability of observing the signal in the data.

Other techniques for estimating signal features exist, for example the RIFT algorithm \cite{lange2018rapid} and DINGO \cite{dax2021real}. For a more detailed explanation of parameter estimation see \cite{thrane2019introduction}.

Possible signals are created through waveform approximations. These waveforms estimate the signal strain in the detector from the set of parameters that the sampler has drawn. The appearance of the waveform depends on the set of parameters that describe the system and the waveform approximation selected. These parameters are divided into two groups, intrinsic and extrinsic.

The intrinsic parameters describe the internal dynamics of the merging binaries. These include the chirp mass, $\mathcal{M}_c$, mass ratio, $q$, and six parameters describing the spins of the compact objects. The extrinsic parameters describe the system relative to the detector. These include the luminosity distance of the system $d_L$, the relative inclination of the orbital plane, sky location in right ascension and declination, the phase, $\phi$, of the wave at arrival, the polarisation of the wave and the time of arrival. This set of 15 parameters must be selected in order to produce possible waveforms for the signal.

For this analysis we used the precessing waveform IMRPhenomPv2 \cite{hannam2014simple, bohe2016phenompv2, husa2016frequency, khan2016frequency} to estimate the parameters of the merging binary. More recent waveforms, as described in ~\cite{garcia2020multimode, pratten2021computationally, khan2020including, ossokine2020multipolar}, allow for more accurate physical descriptions of systems. To follow standard convention we use the IMRPhenomPv2 waveform to keep the majority of this physics without drastically increasing the computational cost.

In PE, spin is often represented in terms of more generic spin parameters. In this paper we will refer to $\chi_p$, the precessing spin parameter. This describes the in-plane spin of the two black holes. See reference \cite{schmidt2015towards} for a more detailed explanation of this parameter. Parameter estimation of BBH mergers with little to no precessing spin generally return a posterior matching the given prior, or a posterior close to zero with small positive deviations. If it is constrained away from zero it is an indication that the spins of the component black holes are likely to be precessing about the orbital angular momentum vector of the system.

Certain parameters can be removed from the sampling by marginalising over the likelihoods of those parameters \cite{creighton2012gravitational}. This reduces the number of parameters in the integral for the likelihood without large uncertainties in the recovered likelihood. In this study the sampling was performed with marginalisations over time, phase and distance. Phase marginalisation is not always possible with precessing waveforms, creating likelihoods that can differ from the true likelihood. However, IMRPhenomPv2 contains both (2,$\pm$2) multipoles in the co-precessing frame \cite{hannam2014simple}. This allows for precession to be measured with marginalisation over phase.

\subsection{Studying overlapping signals} \label{OLsigissues}
In the case of a detected signal, current data analysis techniques assume only the presence of a single signal in the noise. We perform our analysis on a similar basis, in which the data segments containing one or more signals are treated as single events. However, if two or more GWs are present in the detector at the same time their signals will interfere and produce a non-physical composite waveform. The sampling software will then select sets of parameters that match this composite waveform, rather than those of the component signals.

For the scope of this study we restrict overlaps to two signals. The observed waveform is therefore comprised of two component signals, Signal A and Signal B, where Signal A is the primary waveform that remains constant throughout the analysis. We perform parameter estimation on a variety of combinations of the two signals and observe the situations in which the sampler recovers signals that differ significantly from the true posterior of either signal. Throughout our analysis we keep Signal A to be a GW150914-like merger with chirp mass of $28.1\,\mathrm{M_{\odot}}$ and mass ratio $0.806$.

There are four main differences that can describe two overlapping signals. The first being the relative merger time, the time separation between the mergers of Signal A and Signal B. We control this separation in our analysis by manually setting a displacement of Signal B's merger time upon signal creation. For the majority of our analysis, we keep this separation constant between detectors by giving both signals the same sky location. This is of course an unlikely situation, but should have little effect on the outcome of the PE for studies of events in a single detector. For detector networks it does increase the coherence between observed signals in the network, see Secs. \ref{best_case} and \ref{worst_case}. For our primary analysis we allow the merger time of Signal B to vary according to $\Delta t_{\mathrm{merger}}: [-0.1,0.1]\mathrm{seconds}$, relative to the merger time of Signal A.

The second relative parameter compares the SNRs of the two signals, how loud Signal B is in comparison to Signal A. In order to vary this, we keep Signal A at a constant SNR of 30 throughout the whole analysis. The SNR of $30$ was set for the LIGO: Hanford detector. The SNR in LIGO: Livingston was $24.3$. We then vary the luminosity distance of Signal B such that it would independently appear at several, usually lower, SNRs in the detector. For our primary analysis we allowed $\mathrm{SNR_B}: [5,30]$, for the LIGO: Hanford detector, in order to vary from very low relative SNRs to almost equal.

The third relative parameter is the difference in the two signals true waveforms. The frequency evolution of each waveform is described uniquely by its own set of parameters. To this extent we describe the frequency evolution of each signal by controlling its chirp mass, $\mathcal{M}_c$. The chirp mass dominates the first Post Newtonian expansion and as such has a large effect on the frequency evolution of the waveform. To observe how this changes overlaps we vary Signal B's chirp mass according to $\mathcal{M}_c: [24.1,32.1]\,\mathrm{M_{\odot}}$ giving a wide range around that of Signal A. The mass ratio of Signal B was kept constant at 0.9.

The frequency evolution of the two signals is also dominated by their initial phases. For our analysis, we keep Signal A's phase constant and perform several sets of analysis for each configuration of Signal B at different random phases.

We have restricted all analysis to overlapping BBH events. This was done as we expect these types of overlap to cause the most significant bias in parameter estimation within the next decade of ground-based detections, see Sec. \ref{P_BBH_OL} for further information. From preliminary, exploratory, studies we also found that overlapping BBH events could show precession-like effects. This is the case even when the component signals contained little to no precessing spin.

Other studies \cite{pizzati2021bayesian, samajdar2021biases} have considered overlapping BNS events and BBH events that merge during a visible BNS signal. For the sake of reducing computational time we have left these studies for future analysis. However, we expect that BNS+BBH mergers are less likely to cause significant bias due to their vastly different frequency evolutions. Unless the SNR of the BBH signal is very high relative to the BNS, sampling will likely recover the BNS correctly when given priors for the BNS, due to the much larger number of cycles. The signal will remain clean for much of its observable period. If recovery of the BBH is desired then, unless the relative SNR is very large, it is likely that no reliable PE can be performed in this situation.

For the recovery a constant set of priors were used throughout the study. These priors were kept as close as possible to those outlined in Appendix C of \cite{abbott2019gwtc}, in order to best match initial analysis that would be performed in the event of a detection trigger. To fully encompass the parameters of both systems, some of these priors were widened slightly. The luminosity distance prior was set with an upper limit of $6000\,\mathrm{Mpc}$ in order to cover the large luminosity distance of the lowest mass Signal B at SNR 5.

It is possible that the interference of the waveforms could cause merger like effects earlier in the period of the signal. Therefore, the merger time prior was also widened from 0.2 seconds to 1 second. This was done in order to encompass the full chirp of both signals. This would allow for any signals that might interfere in such a way to cause chirp-like characteristics far from the true merger.

The runs were all performed with signals injected in the commonly used "zero-noise" approximation where the noise is assumed to be identically zero across all frequencies. The resultant obtained posteriors are then expected average posterior under a large number of noise realisations. In addition we performed runs with Gaussian noise added to the signal to ensure consistency in our zero-noise results. None of these runs differed qualitatively from the posteriors found in the zero noise case and all conclusions held.

\section{RESULTS} \label{results}
\subsection{Single detector runs} \label{singledetector}
Our primary analysis considered overlapping events as seen in a single detector. Events are rarely published if only found in a single detector, however, this can occur in the event of an interesting event. This is possible for an event which has been significantly biased by an overlapping signal.

For each run; two signals were generated as described in Sec. \ref{method}, Signal B was then injected into the same data frame as Signal A using the python package \texttt{GWpy} \cite{macleod2021gwpy}. The created data frames were then given to the parameter estimation software, \texttt{BILBY}, \cite{ashton2019bilby, romero2020bayesian} using the nested sampler \texttt{Dynesty} \cite{speagle2020dynesty}. The data segments were treated as potential observed BBH events with no further assumptions. Each data segment was 8s long, with 6s prior to the merger time of Signal A. This was done in order to best match the detection process of a true GW event. The PSD used for these runs is the aLIGO Design PSD, from the "Design" column of the table given in \cite{aligoPSD}.

\begin{figure}[t!]
\centering
\includegraphics[width=0.45\textwidth]{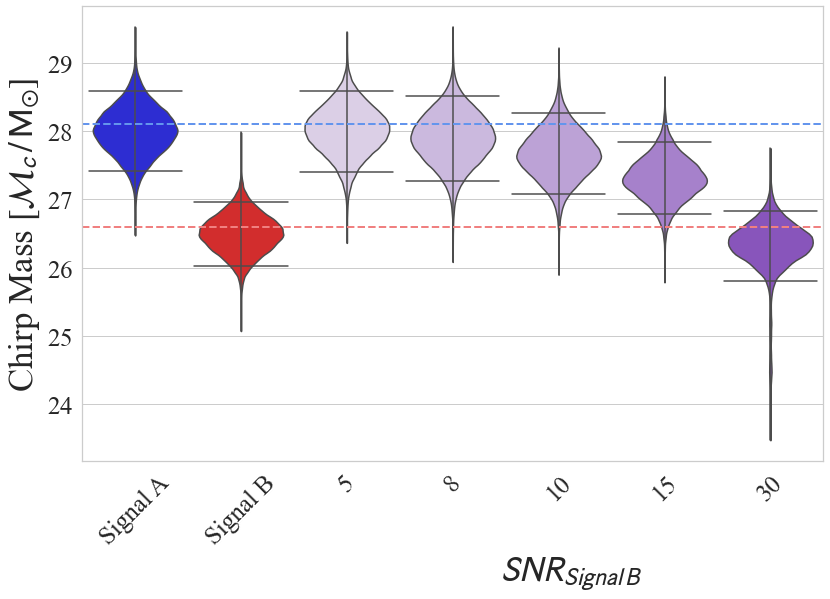}
\caption{Recovered chirp mass posterior distributions. Posteriors labelled Signal A and Signal B are for data with a single signal injected at SNR 30. Signal B is injected with a relative merger time of $+0.025$ seconds and a chirp mass of $26.6\,\mathrm{M_{\odot}}$. The next five posteriors have the two signals injected with the same properties but with the SNR of Signal B varying from 5 to 30. Signal A is kept with an SNR of 30 in all runs. The two horizontal lines in blue and red show the injected values of the chirp mass for Signal A and Signal B respectively.} 
\label{fig:single_SNR}
\end{figure}

We find that in the case of an uneven SNR ratio, ratio of the SNRs of the two signals, the sampler will recover parameters much closer to those of the louder signal. The sampler can almost always correctly recover Signal A in the situation where its SNR is more than three times the SNR of Signal B. The results are more reliable when Signal A shares less of the parameter space with Signal B. If Signal B is allowed to exist at SNRs greater than that of Signal A then the posterior settles on the values of Signal B. The effect of the SNR ratio is shown in Fig. \ref{fig:single_SNR}, where the SNR of Signal B is increased with other signal parameters kept constant.

All plots presented here will be posterior distributions of parameters, in the form of violin plots. The first two posteriors will show the posterior distributions for separate, single signal runs of Signal A and Signal B. The posteriors that follow these are for two overlapping signals of near identical parameters, but with stated differences. Horizontal grey lines show the $90\%$ credible intervals for the posteriors. The two horizontal lines in blue and red show the injected values for the parameter for Signal A and Signal B respectively. All posterior plots were made using the parameter estimation plotting software, \texttt{PESummary} \cite{hoy2021pesummary}.

When the two signals have a time separation greater than the visible duration of the signal in the detector then the sampler will favour the louder signal. In this situation the signals would have been recorded as separate detections and would not be studied with wide enough time priors to cover both signals.

In the situation where Signal A is both louder than Signal B and merges at a later time, the recovered posterior is much closer to that of Signal A. This is because most of the power is within the period of Signal A. However, if Signal B merges after Signal A, the posterior is more likely to recover significant bias, even at lower SNR ratios. The reverse is true if the relative SNRs of the two signals are swapped.

\begin{figure}[t!]
\centering
\includegraphics[width=0.45\textwidth]{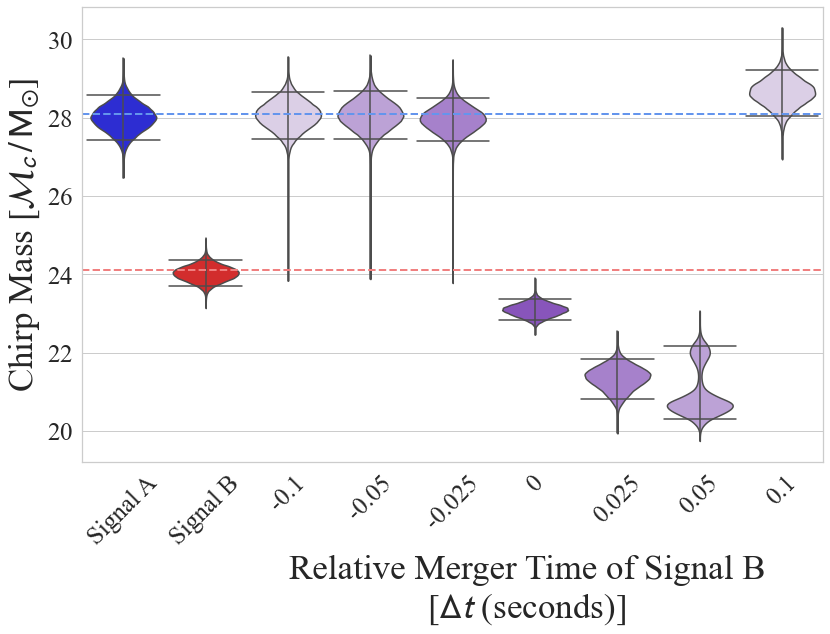}
\caption{Recovered chirp mass posteriors. Posteriors labelled Signal A and Signal B are for data with a single signal injected at SNR 30. Signal B is injected with SNR 30 and a chirp mass of $24.1\,\mathrm{M_{\odot}}$. The next seven posteriors have the two signals injected with the same properties but with the relative merger time of the secondary varying. The two horizontal lines in blue and red show the injected values of the chirp mass for Signal A and Signal B respectively.}
\label{fig:single_time}
\end{figure}

When signals do overlap in the time domain there will always be some bias in the parameter estimation. The bias from time overlap increases as the merger times of the two signals are brought closer together. Fig. \ref{fig:single_time} shows the recovered posteriors for the chirp mass of near identical runs where the SNR of Signal B is constant, but the time separation of the mergers varies between runs.

For highly overlapping signals, the significance of the bias is dominated by the frequency evolution of the signals. The bias is amplified if the signals are close in the frequency domain. If the parameters of the two systems are such that they have a very different frequency profiles, for instance if they have very different chirp masses, then the sampler struggles to build waveform templates that match the wide variations in frequency. The sampler is therefore more likely to match to the louder signal. On the other hand, if the signals have similar frequency evolution then the power in the signals are combined and the sampler has any easy job matching the composite, interfering signal. 

\begin{figure}[t!]
\centering
\includegraphics[width=0.45\textwidth]{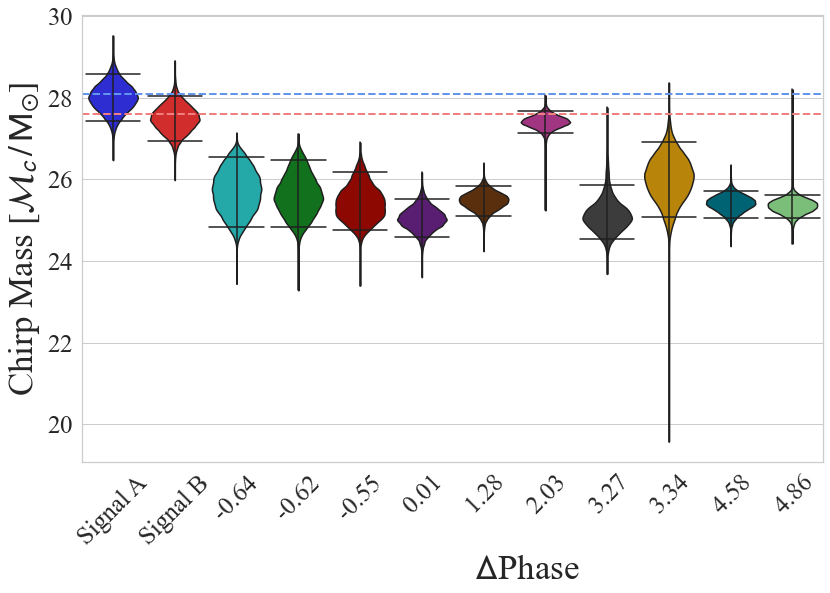}
\caption{Recovered chirp mass posteriors. All runs are identical with Signal B of chirp mass $27.6\,\mathrm{M_{\odot}}$, relative merger time of $-0.025$ seconds and SNR 30. The overlapping signal posteriors differ due to the initial phase of Signal B, which was selected at random from a uniform distribution. The two horizontal lines in blue and red show the injected values of the chirp mass for Signal A and Signal B respectively.}
\label{fig:single_phase}
\end{figure}

The highly overlapping case, in which the two signals have similar SNRs, frequency profiles and merger times, is shown in Fig. \ref{fig:single_phase}. Here there are ten, near identical runs of the same signal configuration. The differences between the recovered posteriors is due to the initial phase of Signal B alone, which was drawn randomly for each run. We performed this analysis for several combinations of Signal B SNR and relative merger time. For each combination we performed the same 10 phase realisations. Fig. \ref{fig:single_phase} shows how much effect the phase has in the highly overlapping case.

\begin{figure}[t!]
\centering
\includegraphics[width=0.45\textwidth]{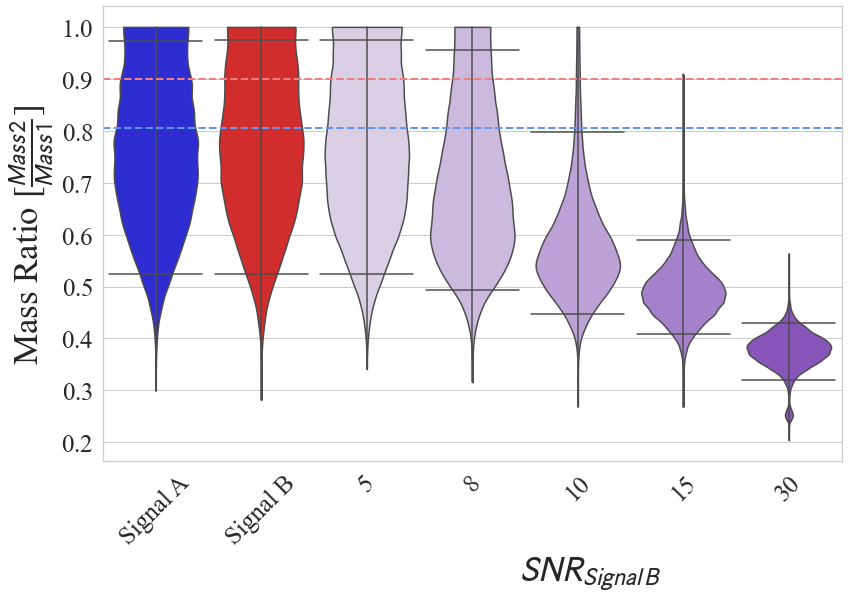}
\caption{A plot of posteriors for recovered mass ratio. Here Signal B has chirp mass $26.6\,\mathrm{M_{\odot}}$ relative merger time of $+0.025$ seconds compared to Signal A. The numerically labelled posteriors are at different injected SNRs of Signal B. Signal A has SNR 30 throughout. The two horizontal lines in blue and red show the injected values of the mass ratio for Signal A and Signal B respectively.}
\label{fig:single_q}
\end{figure}

\begin{figure}[t!]
\centering
\includegraphics[width=0.45\textwidth]{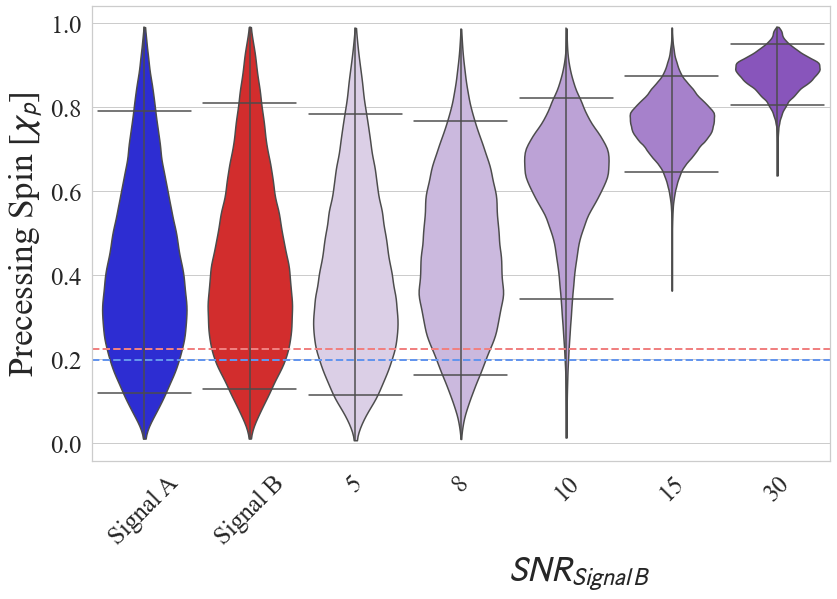}
\caption{A plot of posteriors for recovered $\chi_{p}$. Here Signal B has chirp mass $26.6\,\mathrm{M_{\odot}}$ relative merger time of $+0.025$ seconds compared to Signal A. The numerically labelled posteriors are at different injected SNRs of Signal B. Signal A has SNR 30 throughout. The two horizontal lines in blue and red show the injected values of $\chi_p$ for Signal A and Signal B respectively.}  
\label{fig:single_chi_p}
\end{figure}

In the case of high time-SNR-frequency overlap, the sampler has to fit to a waveform that is varying significantly in frequency space. It therefore, often picks waveforms that are precessing. This leads to samples that have very uneven mass ratios. This is shown in Figs. \ref{fig:single_q} and \ref{fig:single_chi_p}, where the recovered values of mass ratio and $\chi_{p}$ are shown as the SNR of Signal B is increased. The apparent precession caused by overlapping signals is a similar process to one described in \cite{fairhurst2020two}. In this paper they model precession with the beating of two non-precessing signals. Here we have shown that this can occur from two overlapping signals that are visible in the detector.

\subsection{LIGO-Virgo network runs} \label{ligo_network}
Single detector events are rarely published and should always be treated carefully. To examine the more likely case where two overlapping signals are observed in a network of detectors we performed a similar set of injections into three detectors. Two of these detectors were located at the two aLIGO sites in Hanford, Washington and Livingston, Louisiana. These detectors were given an aLIGO design PSD from the penultimate column of the table given in \cite{aligoPSD}. The third detector was located in Pisa, Italy and given a PSD of the Advanced Virgo detector at design sensitivity. This PSD was taken from the penultimate column of the table in \cite{adVirgoPSD}.

In these runs Signal A's parameters were identical to those in the single detector runs, as outlined in Sec. \ref{method}. Signal B was allowed to vary in luminosity distance such that it went from one thirtieth to twice the SNR of Signal A. Runs were performed at a variety of different time separations and phases, but with a single chirp mass for Signal B of $24.1\,\mathrm{M_{\odot}}$.

In multi-detector networks the position of the true signal in the sky controls the time of arrival at each detector. The difference in arrival time between detectors is that of the light travel time between the detectors at a given sky location and arrival time. The maximum difference is between the LIGO: Hanford and the Virgo site in Pisa, this is approximately 27.3 ms. This time separation-sky location dependency means that signals that overlap in the detector interfere differently in different detectors and have different merger time separations in each detector.

Due to this sky location dependency we draw up two cases. The best case, in which the two signals arrive at the detectors from two positions at opposite sides of the line connecting LIGO: Hanford and Virgo. This maximises the difference in arrival time between detectors to 54.6 ms, twice the travel time between them. This situation should lead to the smallest bias causing overlap region and are the most likely overlapping events to be recognised and separated.

We also examine the case in which the two signals arrive from sky locations inducing time delays consistent across the detector network. It is likely that these signals will be more challenging to distinguish from real signals due to the inter-detector consistency. It should be noted that these 'worst case scenario' events are more likely than the best case events. This is because for the merger times to be consistent across the network the signals can either arrive from the same sky location or come from a location perpendicular to the line of sight between detectors. These two scenarios show the extremes of the overlap for detectors based in their current locations.

\subsubsection{Best case scenario} \label{best_case}
We simulate signals at two sky locations at opposite sides of the line of sight between LIGO: Hanford and Virgo. Each signal was created at several time separations, $\Delta t = [-0.1, 0.1]$, as set in the Hanford detector. The resulting Virgo time separations are approximately $\Delta t = [-0.155, 0.045]$. The time separations in LIGO: Livingston also differ slightly from the Hanford values.

For each of these cases four randomly selected values for the phases were given to Signal B. The chirp mass of Signal B was kept to $24.1\,\mathrm{M_{\odot}}$. The same assumptionless, wide prior set and marginalisations were used as in the single detector runs.

In these runs the sampler tends to recover a GW150914-like event for pretty much all the cases in which Signal B merges before the merger of Signal A. In these situations the recovered signal is closer to the parameters of whichever signal is louder in the detector network. This is most likely due to the majority of the power of Signal B being disguised by the more significant Signal A. This is increased by the time separation skewing to before Signal A for the majority of the runs due to the sky location and separation in Virgo.

\begin{figure}[t!]
\centering
\includegraphics[width=0.45\textwidth]{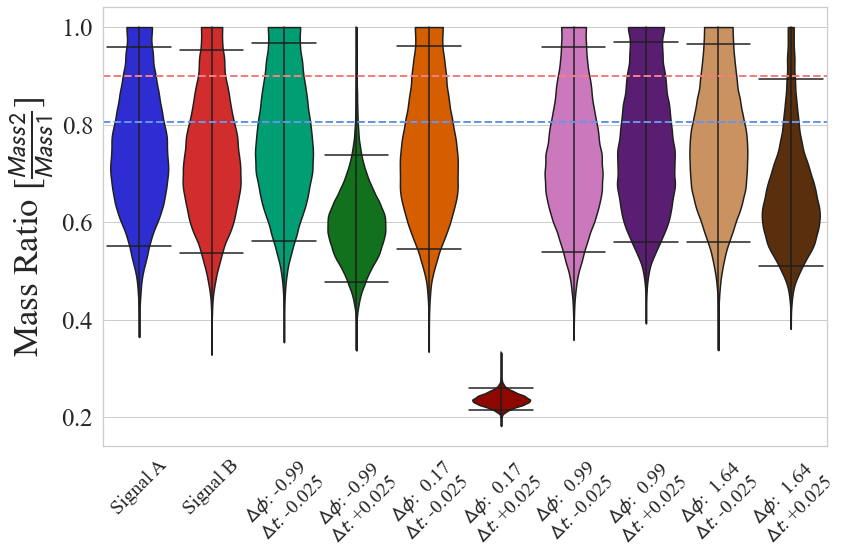}
\caption{Recovered mass ratio posteriors for Signal B overlapping either $0.025$ seconds before or after Signal A at a variety of initial phases, injected into a LIGO-Virgo network. Signal B was injected with an SNR of 25. The sky locations of the two systems of Signal A and Signal B were selected to maximise the travel time between the LIGO: Hanford and Virgo detectors. Stated relative merger times apply to the Hanford detector. The two horizontal lines in blue and red show the injected values of the mass ratio for Signal A and Signal B respectively.}
\label{fig:network_q_best}
\end{figure}

\begin{figure}[t!]
\centering
\includegraphics[width=0.45\textwidth]{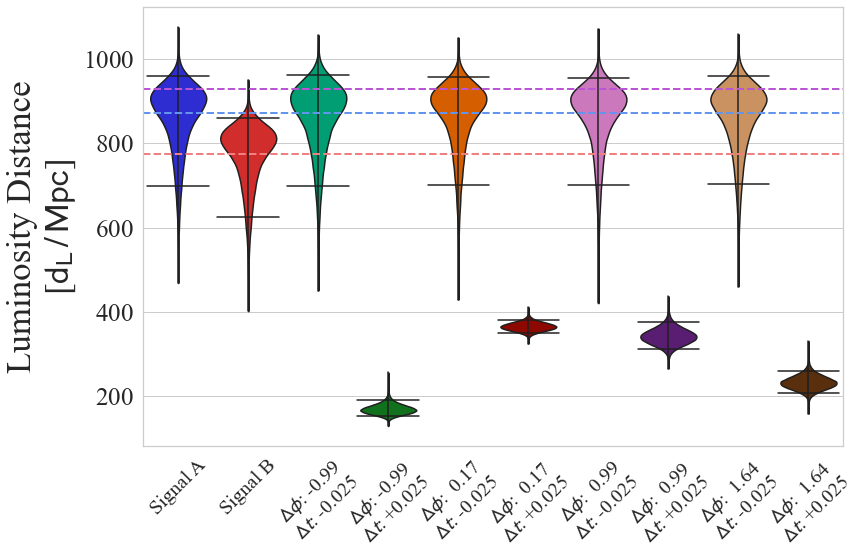}
\caption{Recovered luminosity distance posteriors for Signal B overlapping either $0.025$ seconds before or after Signal A at a variety of initial phases, injected into a LIGO-Virgo network. Signal B was injected with an SNR of 25. The sky locations of the two systems of Signal A and Signal B were selected to maximise the travel time between the LIGO: Hanford and Virgo detectors. Stated relative merger times apply to the Hanford detector. The two horizontal lines in blue and red show the injected values of the luminosity distance for Signal A and Signal B respectively. The final horizontal line shows the injected luminosity distance for Signal B in the combined runs at the lower SNR.}
\label{fig:network_LD_best}
\end{figure}

The recovered posteriors show most evidence of bias in overlapping situations where Signal B merges after Signal A. This can be seen in Figs. \ref{fig:network_q_best} and \ref{fig:network_LD_best}. These show the recovered mass ratio and luminosity distance posterior distributions for the case where Signal B has an SNR of 25 and merges at time separations of $\pm 0.025$ seconds, relative to Signal A. It can be seen that, in the case where Signal B merges $0.025$ seconds after Signal A, in the LIGO: Hanford detector, the bias is much larger than for the equivalent case of earlier merger. At wider time separations the recovery skews toward a GW150914-like event, although significant bias still occurs in some events.

For the case of $\mathrm{SNR}_{A} > \mathrm{SNR}_{B}$; if Signal B merges before Signal A then the majority of the power of Signal B is contained within the inspiral of Signal A. The composite waveform does not differ significantly from that of Signal A. However, if Signal B merges after Signal A, the composite waveform appears Signal A-like, but with a second chirp or elongated chirp. Therefore the sampler tries to match to this composite waveform, selecting more exotic sets of parameters.

However, we find that in all of these runs there is still significant bias in runs that merge with a small time separation. As with the single detector analysis, the bias is reduced at wider time separations and at less even SNR ratios. Effects such as recovered precession and well constrained values of mass ratio, as found in single detector runs, are also present in these highly overlapping network events. 

As in the single detector runs the recovered signal appears highly precessing. With mass ratio and $\chi_{p}$ constrained away from 1 and 0 respectively. To account for this the sampler selects waveforms that look like an edge on system. The sampler therefore predicts that all the power of the signal is in the plus polarisation and leaves little in the cross polarisation. The luminosity distance is therefore recovered to be much smaller than the real value, in order to account for the high SNR. This is shown in Fig. \ref{fig:network_LD_best}.

\subsubsection{Worst case scenario} \label{worst_case}
We performed runs in which the two signals merge at the same location in the sky, relative to the detectors. This situation is the most likely to cause a signal that is not recognised as overlapping due to the relative merger time remaining constant regardless of detector. The relative merger times of these signals are constant.

\begin{figure}[t!]
\centering
\includegraphics[width=0.45\textwidth]{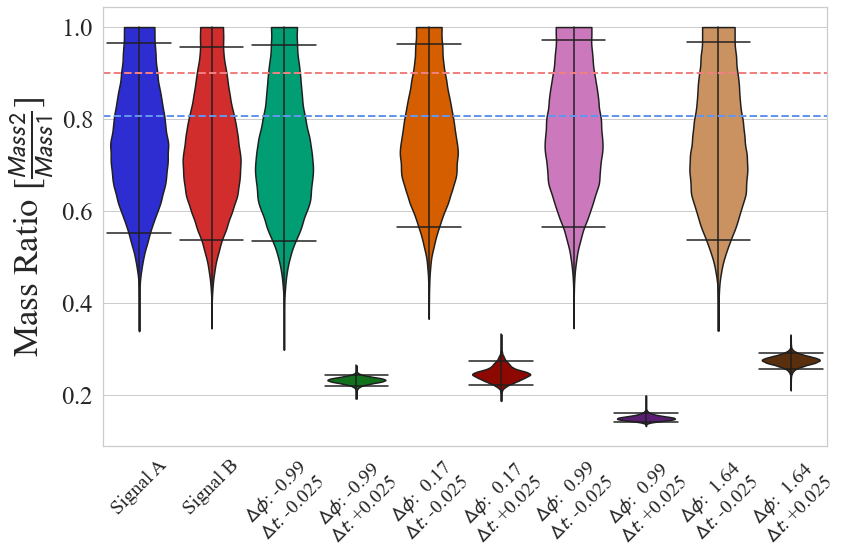}
\caption{Recovered mass ratio posteriors for Signal B overlapping either $0.025$ seconds before or after Signal A at a variety of initial phases, injected into a LIGO-Virgo network. Signal B was injected with an SNR of 25. The sky locations of the two events are identical creating identical arrival times at all three detectors. The two horizontal lines in blue and red show the injected values of the mass ratio for Signal A and Signal B respectively.}
\label{fig:network_q_worst}
\end{figure}

\begin{figure}[t!]
\centering
\includegraphics[width=0.45\textwidth]{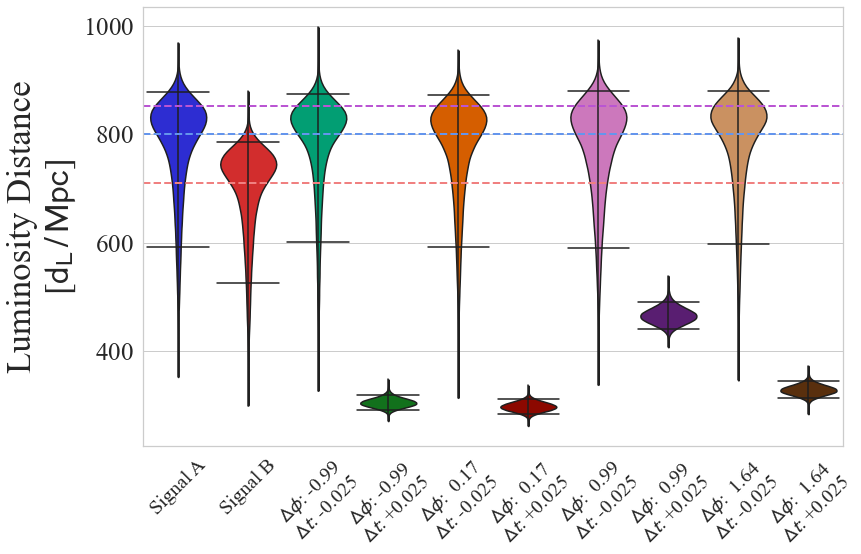}
\caption{Recovered luminosity distance posteriors for Signal B overlapping either $0.025$ seconds before or after Signal A at a variety of initial phases, injected into a LIGO-Virgo network. Signal B was injected with an SNR of 25. The sky locations of the two events are identical creating identical arrival times at all three detectors. The two horizontal lines in blue and red show the injected values of the luminosity distance for Signal A and Signal B respectively. The final horizontal line, purple, shows the injected luminosity distance for Signal B in the combined runs at the lower SNR.}
\label{fig:network_LD_worst}
\end{figure}

Here we find similar results to those in the Sec. \ref{best_case}, however, we find that the observed bias is much larger and skews away from the recovery of a GW150914-like event, particularly in the situation where Signal B merges after Signal A, relative to the detectors. The higher significance of the bias is due to the coherence of the composite waveforms across the network. The two signals interfere in the same manner in all detectors due to their identical sky location. These results are shown in Figs. \ref{fig:network_q_worst} and \ref{fig:network_LD_worst}. 

Similar runs were performed with only the two aLIGO detectors at design sensitivity. The posterior distributions of these runs had the same qualitative results but with less precise constraints on certain parameters, as expected with fewer detectors.

\subsection{Events below the detection threshold} \label{low_snr}
It is possible that signals can be biased by the presence of a second signal, even if that signal were not detectable itself. To test this we performed PE on systems where the Signal A was identical to the GW150914-like signal described previously, but with an SNR of 8 in the LIGO: Hanford detector. This is often regarded as the threshold SNR for signal detection in an individual detector \cite{abbott2020prospects}.

In this analysis Signal B merged $0.025$ seconds after Signal A in order to provide significant bias. Runs were performed with increasing SNR for Signal B from $1$ to $8$. These runs were performed for the full, three detector LIGO-Virgo network at design sensitivity. The network SNR of Signal A was 13.3, slightly higher than the network SNR threshold of 12 \cite{abbott2020prospects}. Sky locations were defined such as to give the two signals a consistent relative merger time across the network of detectors.

\begin{figure}[t!]
\centering
\includegraphics[width=0.45\textwidth]{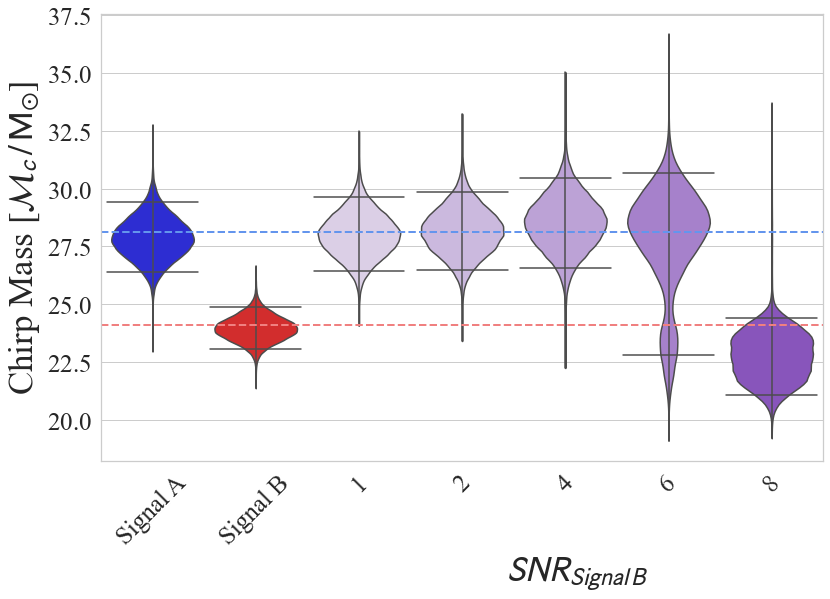}
\caption{Recovered chirp mass posteriors for Signal B overlapping either $0.025$ seconds after Signal A at a variety of different SNRs, injected into a LIGO-Virgo network. The values given for each posterior are the SNR of Signal B in the LIGO: Hanford detector. Signal A was injected with an SNR of 8. The two horizontal lines in blue and red show the injected values of the chirp mass for Signal A and Signal B respectively.}
\label{fig:snr_threshold}
\end{figure}

The posterior distributions for the chirp mass of these runs are given in Fig. \ref{fig:snr_threshold}. These posteriors show that, while these signals are not detectable, they can still cause significant bias in the recovery of Signal A when approaching equal SNR. Such signals would also cause bias in the recovery of a louder Signal A, however, this would not be as significant an issue due to the largely uneven ratio of SNRs.

\subsection{LIGO-Voyager network runs} \label{voyager_network}
Identical runs to the three detector aLIGO network, worst case scenario, were performed with a LIGO-Voyager detector sensitivity. The network kept the same locations as the two aLIGO detectors in Hanford and Livingston, with identical signal parameters. A third detector at LIGO-Voyager sensitivity was not included in the location of the Virgo detector as no current plans for such a detector were available.

Two sets of runs, identical to those described in \ref{ligo_network}, were performed. These runs were performed with two differences. The aLIGO design PSD was replaced with a predicted LIGO-Voyager PSD. The PSD used was created from an estimate of the noise budget calculated through the python package \texttt{pygwinc} \cite{pygwinc}.

One set of runs kept the low frequency cutoff equal to that of the aLIGO network runs, $20\,\mathrm{Hz}$, the other reduced this to $10\,\mathrm{Hz}$, as expected for LIGO-Voyager. The increase in detector sensitivity, combined with identical parameters to the aLIGO network runs, produces signals with much greater significance in the LIGO-Voyager detector. However, it is the relative SNR that is relevant in these situations. This remains constant.

\begin{figure}[t!]
\centering
\includegraphics[width=0.45\textwidth]{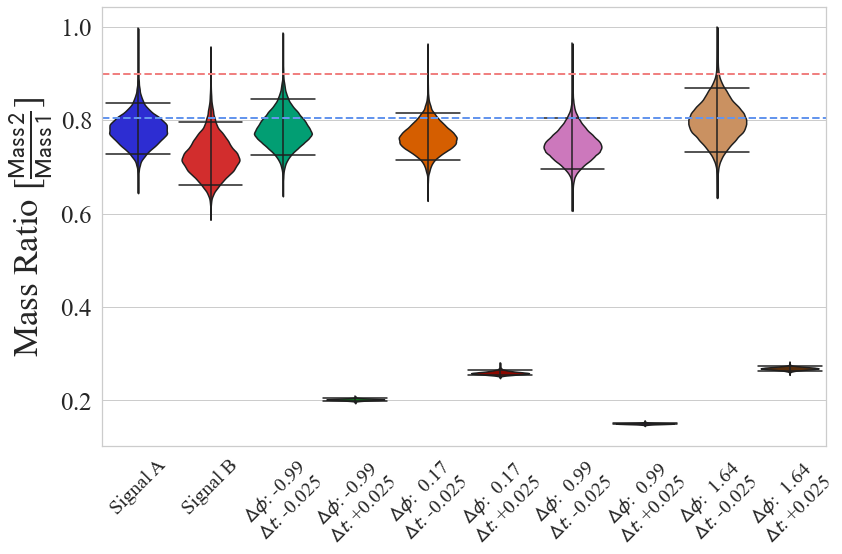}
\caption{Recovered mass ratio posteriors for Signal B overlapping either $0.025$ seconds before or after Signal A at a variety of initial phases, injected into a LIGO-Voyager network. These plots are for the equivalent runs in the aLIGO network, given in Fig. \ref{fig:network_q_worst}. The two horizontal lines in blue and red show the injected values of the mass ratio for Signal A and Signal B respectively.}
\label{fig:voy_q}
\end{figure}

The results of these runs did not differ greatly from those of the aLIGO network. The recovered posteriors were much more precise than from the aLIGO runs, particularly for the mass ratio, Fig. \ref{fig:voy_q}. Better constraints of the posteriors were found in the $10\,\mathrm{Hz}$ runs than the $20\,\mathrm{Hz}$ runs. This is due to the increased number of cycles visible in the detector. This increased number of cycles also shows more interference between overlapping signals. As such the $10\,\mathrm{Hz}$ runs show more bias from overlapping signals than the $20\,\mathrm{Hz}$ runs in all scenarios.

\begin{figure}[t!]
\centering
\includegraphics[width=0.45\textwidth]{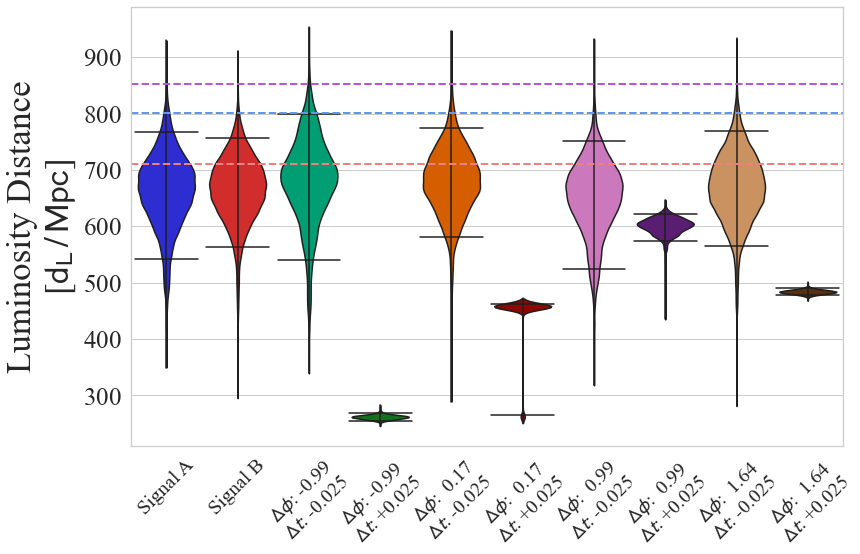}
\caption{Recovered luminosity distance posteriors for Signal B overlapping either $0.025$ seconds before or after Signal A at a variety of initial phases, injected into a LIGO-Voyager network.These plots are for the equivalent runs in the aLIGO network, given in Fig. \ref{fig:network_LD_worst}. The two horizontal lines in blue and red show the injected values of the luminosity distance for Signal A and Signal B respectively.}
\label{fig:voy_LD}
\end{figure}

One notable difference between these runs and those in Sec. \ref{worst_case} can be seen by comparing Figs. \ref{fig:network_LD_worst} and \ref{fig:voy_LD}. The distance posteriors shown in Fig. \ref{fig:voy_LD} are much broader than those in Fig. \ref{fig:network_LD_worst} and biased towards smaller distances. This is an effect of the position of the detectors. The Voyager network is a two detector network and therefore cannot provide as accurate a location of the source as the three detector network, despite the higher sensitivity of the Voyager detectors. The wider sky localisation leads to poorer constraints on distance and a favouring of nearer sources.

Example posteriors of the $10\,\mathrm{Hz}$ runs are shown in Figs. \ref{fig:voy_q} and \ref{fig:voy_LD}. These show identical runs from the $10\,\mathrm{Hz}$ LIGO-Voyager runs as are show in Figs. \ref{fig:network_q_worst} and \ref{fig:network_LD_worst} for the aLIGO network, worst case scenario.

\section{PROBABILITY RE-ESTIMATION} \label{upprob}
Our original investigation into the probability of overlapping signals shows that these events are unlikely to occur until the end of the second generation of GW interferometers. We find that, in a years observations with LIGO-Voyager, there should be approximately $5.3^{+8.6}_{-3.1}$ and $11.0^{+60.0}_{-10.0}$ overlapping BBH and BNS events respectively. There will also be around $230.0^{+360.0}_{-130.0}$ BBH events that merge during the observable period of a BNS merger. However, our study into parameter estimation of such events indicates that the signals need to be within approximately $\pm0.1$ seconds in merger times in order to cause significant bias. Here we consider what effect this has on the probability of observing overlapping signal bias.

We follow a similar analytical process to that outlined in \ref{probOL}. However, to enforce the bias-time difference constraint we shorten the periods of the signals to 0.2 seconds in length. When applying this constraint LIGO-Voyager becomes unlikely to see significant bias causing overlaps within a years observation. When applied to Einstein Telescope there are still some events that will show significant bias. For a years observations this is expected to be approximately: $1.11^{+1.82}_{-0.66}$, $1.32^{+7.07}_{-1.24}$, $1.11^{+1.82}_{-0.66}$ for overlapping BBH, BNS and BNS+BBH\footnote{It should be noted that this number also applies to the number of BBH events that overlap BNS mergers, as both signals have an overlap-bias period of $\sim0.2$ seconds} respectively, for a year of observations. This accounts for less than $0.001\%$ of BBH mergers. While this is unlikely to significantly bias broader population studies via incorrect parameter limits, an interesting looking event could lead to inaccuracies in understanding of precession-like effects.

Despite this, bias can still occur from overlapping signals with relative merger times $> \pm 0.1$ seconds. This wont be as severe as shown in some of the highly overlapping results, but it could still be quite different from the true astrophysical source.

These values are only valid for two signals overlapping the signal. We find negligible probability of three or more signals overlapping within $\sim0.2$ seconds, even in third generation detectors. However, our analysis of the significant bias region only applies to two signal overlaps. It is likely that multiple signals polluting the data will cause significant deviations from the true waveforms and that current parameter estimation techniques will struggle to account for this.

\section{CONCLUSION AND FUTURE WORK} \label{conclusions}
Our analysis shows that, although we are unlikely to observe overlapping signals within the lifetime of current detectors, we should expect such events to occur in observations in LIGO-Voyager. This is most likely to occur in the case of BNS+BBH overlaps. However, we do not expect these events to produce significant bias due to the significant differences in their frequency evolution. Unless the merger times are within $\sim0.1$ seconds of each other; the sampler will recover a largely unbiased signal. This signal should be one which best fits the given mass prior. Further analysis should be performed to confirm this.

We find that overlapping signals will be present in small numbers by the end of the second generation with LIGO-Voyager. We also find that third generation detectors, such as Einstein Telescope, will be so sensitive that they are unlikely to ever observe signals that do not overlap with multiple other signals. However, due to the requirement of close merger times and high relative SNR, many signals should remain recoverable to a reasonable approximation.

We also predict that LIGO-Voyager should detect several overlapping BBH events. While a few, low overlap, cases may not be a significant problem, their presence could cause trouble for studies of binary black hole populations and tests of general relativity. In the case of detecting a highly overlapping event; the system could appear highly precessing leading to a further detailed analysis and publication. These events may affect any further astrophysical studies. We have only considered this for single event cases and leave the effects on population studies for future works.

Overall we find that overlapping BBH mergers do not show significant bias unless their waveforms overlap significantly in time and frequency. The ratio of the SNRs of the two signals must also be fairly even, as high as 0.3, for the sampler to recover signals that differ significantly from the true posterior of the louder signal. Generally bias is more significant if the quieter signal merges after the primary signal.

It is reasonable to assume that most early detections of overlapping BBH mergers will not show significant bias. They should, however, only recover the louder signal. To recover the quieter signal new analysis techniques must be found to sample over the parameter space of two signals. Recently, some methods have been proposed that may solve this problem ~\cite{smith2021bayesian, antonelli2021noisy}.

In a network of detectors the relative sky locations of the two signals, and their relation to the line of sight between the detectors, has an effect upon the significance of the bias. We find that signals in sky locations along the line of sight between two detectors have a smaller region of significant bias. However, if the signals are in locations perpendicular to the line of sight, or at similar sky locations, the bias is more significant. 

In the situation of two highly overlapping waveforms the position and distribution of the posterior varies wildly with respect to the initial phase of the two waveforms. In these scenarios the sampler often selects waveforms that match highly precessing signals. This is due to the modulation of the frequencies present in the combined waveform. The result is often systems with mass ratios well constrained and away from the even case. The sampler will then put the majority of the power into the cross polarisation and compensate by constraining a low luminosity distance.

We find that biases from overlapping signals can occur even if the quieter signal causing the bias is itself undetectable. Detections of events with low SNRs should be treated with caution due to the possibility of bias from an undetectable signal.

The results for LIGO-Voyager show that, while higher-sensitivity detectors are more likely to detect bias, they are also more likely to present significant biases in detections. If these can be properly characterised it should be possible to establish which detections contain a potential overlap. It may be necessary to look closely at any events that show significant evidence of precession or uneven mass ratios.

Our analysis of the effects of overlapping waveforms has been confined to the case of overlapping BBH mergers. We don't expect to observe the overlap of BNS systems until significant improvements are made in low-frequency detector sensitivity. This limits the observation of such overlaps until third generation detectors.

We did not examine the case of a BBH signal overlapping the merger of a BNS signal. However, we expect that the mergers of BBH events that occur within the inspiral of BNS events should not be recoverable unless they are much louder than the BNS signal. In this case a process similar to the gating of glitches can be used to remove the BBH for analysis of the BNS signal. When the BNS is louder than the BBH then we would expect only the BNS to be recoverable.

For the less likely case that a BNS merger is close to a BBH merger we expect the signals to not be easily recoverable without bias. It may be possible to correctly recover the BNS from its inspiral. Further, it may  be possible to use the inspiral to predict the merger of the BNS. In this case it can be removed from the data to reveal the BBH merger for further analysis. This could be a viable method of distinguishing such events and we leave this for future analysis. However, we have not performed any analysis on overlapping BNS systems.

We do not present any complete methods for proving that a detected signal is biased by another. Neither do we present methods of sampling these situations to distinguish between the signals. However, we have expect that it may be possible to do so by comparing the appearance of the waveform between the two detectors and examining the cases of highly precessing, highly polarised signals.

In the mid 2030s a space based Gravitational Wave interferometer is planned to be launched. The Laser Interferometer Space Antenna, LISA, detector \cite{amaro2017laser} is planned to have a much lower sensitivity than aLIGO. However, the detector is designed to observe a different frequency range than ground based detectors. Sensitive between the $10^{-5}\,\mathrm{Hz}$ and $10^{-1}\,\mathrm{Hz}$ band, it should observe the mergers of much more massive objects than LIGO. These more massive signals, such as supermassive black hole mergers, will be much slower than LIGO BBH mergers. It will not be possible to study these signals without accounting for their interference from signal overlap.

\section{ACKNOWLEDGEMENTS} \label{acknowledgements}
We would like to thank Charlie Hoy and Duncan Macleod for invaluable programming assistance. We would also like to thank Fabio Antonini, Sebastian Khan and Geraint Pratten for useful discussions. Further we would like to thank the anonymous reviewers for their constructive discussions. We are grateful for computational resources provided by Cardiff University, and funded by an STFC grant supporting UK Involvement in the Operation of Advanced LIGO, ST/V001337/1. This work was supported in part by STFC grant Nos. ST/V001396/1, ST/S505328/1 and ST/V00154X/1. The work has been allocated the internal LIGO Document number LIGO-P2100074.

\bibliographystyle{unsrt}
\bibliography{main.bib}
\end{document}